\documentclass[aps,pre,reprint,superscriptaddress,showpacs,preprintnumbers,amsmath,amssymb,showkeys]{revtex4-1}
\usepackage{amssymb}
\usepackage{amsmath}
\usepackage{graphicx}
\usepackage{color}
\usepackage{soul}
\usepackage{float}
\usepackage[english]{babel}

\begin{document}

\title{Temporal Pattern of Online Communication Spike Trains in Spreading a Scientific Rumor: How Often, Who Interacts with Whom?}



\author{Ceyda Sanli}
\email{cedaysan@gmail.com}
\affiliation{CompleXity and Networks, naXys, Department of Mathematics, University of Namur, 5000 Namur, Belgium}

\author{Renaud Lambiotte}
\affiliation{CompleXity and Networks, naXys, Department of Mathematics, University of Namur, 5000 Namur, Belgium}

\date{\today}

\begin{abstract}

We study complex time series (spike trains) of online user communication while spreading messages about the discovery of the Higgs boson in Twitter. We focus on online social interactions among users such as retweet, mention, and reply, and construct different types of active (performing an action) and passive (receiving an action) spike trains for each user. The spike trains are analyzed by means of local variation, to quantify the temporal behavior of active and passive users, as a function of their activity and popularity. We show that the active spike trains are bursty, independently of their activation frequency. For passive spike trains, in contrast, the local variation of popular users presents uncorrelated (Poisson random) dynamics. We further characterize the correlations of the local variation in different interactions. We obtain high values of correlation, and thus consistent temporal behavior, between retweets and mentions, but only for popular users, indicating that creating online attention suggests an alignment in the dynamics of the two interactions.

\end{abstract}


\maketitle

\section{Introduction}
\label{intro}

%
%

In recent years, online social media (OSM) have become a major communication channel, allowing users to share information in their social and professional circles, to discover relevant information pre-filtered by other users, and to chat with their acquaintances. In addition to their practical use for individuals, OSM have the advantage of generating a rich data set on collective social dynamics, as social relations among individuals, temporal properties of their interactions, and their contents are automatically stored. The study of these digital footprints has led to the emergence of computational social science, allowing to quantify at large-scales our political ideas and preferences~\cite{Abisheva:2014:WYU:2556195.2566588}, to discover roles in social network~\cite{Wu:2011:SWT:1963405.1963504, 6542426}, to predict our health~\cite{predictingdepression} and personality~\cite{perso}, and to determine external effects on online behavior~\cite{Mathiesen22102013}. Importantly, in OSM, users are at the same time both actors and receivers and therefore the amplification of a trend originates from the interplay between influencing~\cite{Bakshy:2011:EIQ:1935826.1935845, broadcastershiddeninfluencers} and being influenced~\cite{Bakshy:2009:SID:1566374.1566421, MoreVoicesThanOthers, SNIOnlineBehaviorChoices, Zhang:2015:IYP:2737800.2700398, 2015arXiv150603022L}.

A crucial aspect of OSM and more generally of human behavior is the underlying complex dynamics~\cite{BarbarasiOriginofBursts, 0295-5075-81-4-48002, CorrelationCommuication, PhysRevE.90.012817}. The time series of user activities, e.g. posting a tweet and replying to a message, are quite distinct from uncorrelated (Poisson random) dynamics in the presence of burstiness~\cite{0295-5075-94-1-18005, 0295-5075-97-1-18006, PhysRevE.92.012817}, temporal correlations~\cite{Mathiesen22102013, UniversalCorrelatedBursts, PhysRevE.92.022814}, and non-stationarity of human daily rhythm~\cite{Circadianpatternburstiness_mobilephone, 10.1371/journal.pone.0058292}, which has significant implications. Diffusion on a temporal network cannot be accurately described by models on static networks and consequently the process presents non-Markovian features with strong influence on the time required to explore the system~\cite{PhysRevLett.103.038702, LambiotteTabourierEPJB_Bursts}. Furthermore, the dynamics drives a strong heterogeneity observed in user activity~\cite{ HowRandomOnlineSocialInt, 2014arXiv1405.5726F} and user/content popularity ~\cite{Ferrara:2014:OPT:2631775.2631808, onlinepopularityheterogeneity, Competition_memes}. Specifically, in Twitter, the heterogeneity in popularity has been observed and quantified in different ways by the size of retweet cascades, i.e. users re-transfer messages to their own followers with or without modifying them~\cite{burstydynamicsTwitter, TrendsinRetweetDynamics, SEISMIC, 5428313, export:141866} or by the number of mentions of a user name, identified by the symbol $@$, in other people's tweets~\cite{mentionnetworkTwitter}. 

\begin{figure}[h!]
\begin{center}
\includegraphics[width=8.7cm]{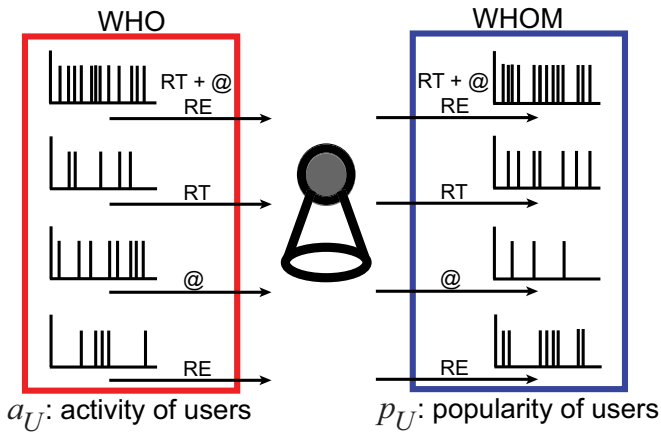}
\end{center}
\vspace{-6mm} \caption{ \label{fig:01} \small Illustration of communication in Twitter. Users in who interact in time with users in whom by retweeting (RT) the messages and mentioning (@) the user names of whom in a message and replying (RE) to the messages from whom. Quantifying temporal patterns in time series of who users with various ranges of the activity of users $a_U$ and of whom users by increasing the popularity of users $p_U$ is the main scope of this paper.}
\end{figure}

In this paper, we focus on the dynamics of social interactions taking place when diffusing rumors about the discovery of the Higgs boson on July 2012 in Twitter~\cite{scientificrumor}. Our main goal is to find connections between the statistical properties of user time series established on the same subject, e.g. the announcement of the discovery of the Higgs boson, and their activity and popularity. To this end, we analyze tweets including social interactions, such as retweets of a message (RT), mentions of a user name (@), and replies to a message (RE). For each type of the interactions, a user can either play an active, e.g. retweeting, or a passive, e.g. being retweeted, role. Therefore, we characterize each user by 8 time series: one active and one passive time series for each of the 3 types of interaction as well as for the aggregation of all interactions, as illustrated in Fig.~\ref{fig:01}. Active time series are denoted as WHO and passive time series are defined by WHOM. We then investigate whether the statistical properties of each signal is a good predictor for the activity and popularity of a user.

The following sections are organized as follows. In section~\ref{sec:aUpU}, we describe the data set and provide basic statistical properties of who and whom time series. In section~\ref{sec:Lv}, we introduce a technique dedicated to the analysis of non-stationary time series, so-called local variation, originally established for neuron spike trains~\cite{Lv1, Lv3, Lv5, LVcorrelation} and recently has been applied to hashtag spike trains in Twitter~\cite{10.1371/journal.pone.0131704, LvCollectiveAttention}. In section~\ref{sec:TP}, we search for statistical relations between local variation and measures of popularity of a user. Finally, section~\ref{sec:Discussion} summarizes the key results and raises open questions.

\section{Activity and Popularity of Users}
\label{sec:aUpU}

Our aim is to examine the dynamics of user communication in Twitter. We  investigate how frequently Twitter users talk to each other on a certain topic, e.g. the discovery of the Higgs boson, and identify how complex dynamic patterns of the communication evolve in time. To this end, we focus on the 3 different types of interaction between users, retweet (RT), mention (@), and reply (RE). Twitter users can adopt a tweet of someone and use it again in their own tweet blog by RT or contact to other users directly by typing user names in a message called @ or simply RE to any tweets, e.g. regular tweets, retweets, and tweets/retweets including @s. Typically, @s and REs are associated to personal interactions between users, whereas RTs are responsible for large-scale information diffusion in the social network and present cascades. Here, we count all types of interaction as a part of complex information diffusion in Twitter.

Interactions in Twitter are performed between at least two users (for instance, a user can mention several other users in a single tweet). Each action is directed and characterized by its timestamp. The users performing the action play active roles (who users) and the users receiving their attention play a passive role (whom users). We construct active and passive RT, @, and RE spike trains for each user. 

\textit{\textbf{Data Set.}} 
As a test bed, we consider the publicly available Higgs Twitter data set~\cite{scientificrumor, snapnets}, first collected to track the spread of the rumor on the discovery of the Higgs boson via RT, @ or RE. The data set is composed of tweets containing one of the following keywords or hashtags related to the discovery of the Higgs boson,  ``lhc", ``cern", ``boson", and ``higgs". The start date is the 1st July 2012, 00:00 am and the final date is the 7th July 2012, 11:59 pm, which covers the announcement date of the discovery, the 4th July 2012, 08:00 am. All dates and timestamps in the data are converted to the Greenwich mean time. Detailed information on the data collection procedure and basic statistics can be found in Ref.~\cite{scientificrumor}.

\begin{figure}[h!]
\begin{center}
\includegraphics[width=8.3cm]{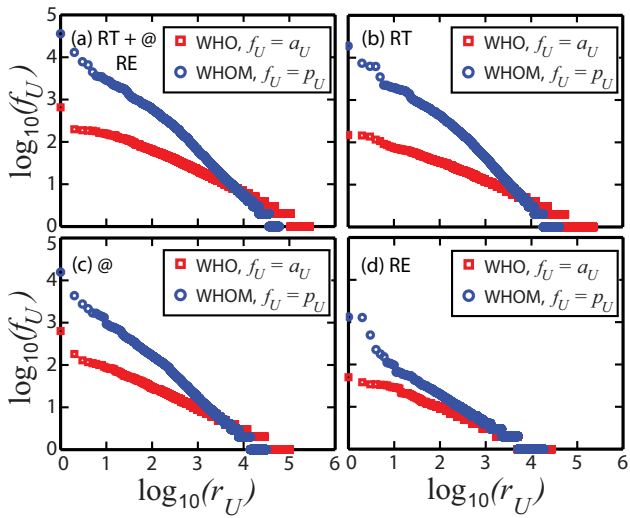}
\end{center}
\vspace{-6mm} \caption{ \label{fig:02} \small How often who users communicate with whom users. Zipf plots describe heterogeneities of users in the types of Twitter interaction, e.g. (a) all of retweet, RT, mention, @, and reply, RE, (b) only RT, (c) only @, and finally (d) only RE. The frequency of the communication $f_U$ is measured in two-fold: The activity of who users (red squares) $a_U$ and the popularity of whom users (blue circles) $p_U$. The $x-$axis ranks the users $r_U$ from high $f_U$ to low values. Each plot indicates that Twitter users, who users, more likely contact to someone observed in smoother decays of who users, however only few whom users are addressed and become popular in these communications.}
\end{figure}

In total, the data is composed of 456,631 users (nodes) and 563,069 interactions. Among those, we detect 354,930 RT, 171,237 @, and 36,902 RE, which shows that RT is more popular than the other communication channels. For RT interactions, we find  228,560 who and 41,400 whom users. These numbers are smaller for @, e.g. 102,802 who and 31,477 whom users, and even smaller for RE, with 27,227 who and 18,578 whom users. In each case, whom is much lower than who, as expected because a small number of users tend to attract a large fraction of attention in both friendship~\cite{Saramaki21012014, Meo:2014:AUB:2542182.2535526} and online social~\cite{Twitterundermicroscope, 10.1371/journal.pone.0022656, Competition_memes_limitedattention, Competition_inducedcriticality, Competition_advertisement} networks. This observation is confirmed in Fig.~\ref{fig:02}, where we present the Zipf plots associated to each interaction, clearly showing a strong heterogeneity in the system. For who users, the frequency of the user communication $f_U$ ranks how active users are and measures the activity of users $a_U$, on the other hand, for whom users, $f_U$ quantifies how often the users or their tweets are addressed and so gives the popularity of users $p_U$ .

\section{Local Variation of Who and Whom}
\label{sec:Lv}

\textit{\textbf{Communication Spike Trains.}} We extract salient temporal patterns of the user communication time series. We evaluate each directed interaction (RT, @, and RE) of the users in the pool of who with any users in the whom class, shown in Fig.~\ref{fig:01}. We don't check whether the whom users participate the conversation in a later stage and only construct independent time series of the who and whom users. The elements of the time series are the timestamps of the data~\cite{scientificrumor, snapnets} providing us the exact time in second of the interaction and the user name or ID of the corresponding who and whom users. Ordering the timestamps from the earliest to the latest, we generate spike trains carrying full story of the communication of each user. The resultant user communication spike trains are grouped in eight: For each who and whom, the spike trains of all interactions together (i) and the spike trains of filtered timestamps of RT (ii), @ (iii), and RE (iv).   

\textit{\textbf{Local Variation.}} A standard way of investigating the dynamics of human communication is to examine the statistics of the inter-event spike intervals such as its probability distribution~\cite{BarbarasiOriginofBursts}, short-range memory coefficient and burstiness parameter~\cite{0295-5075-81-4-48002} or Fano factor. However, recent works have showed that further detail analysis is required to resolve temporal correlations~\cite{Competition_memes, burstydynamicsTwitter}, bursts~\cite{0295-5075-97-1-18006, PhysRevE.92.012817, UniversalCorrelatedBursts, PhysRevE.92.022814}, and cascading~\cite{CascadingBehavior} driven by circadian rhythm~\cite{Circadianpatternburstiness_mobilephone, 10.1371/journal.pone.0058292}, complex decision-making of individuals~\cite{6542426, HowRandomOnlineSocialInt, PhysRevE.85.066101}, and external factors~\cite{Mathiesen22102013} such as the announcement of discoveries, as considered in the current data~\cite{scientificrumor}.

To uncover the dynamics of the communication spike trains elaborately, we apply the local variation $L_V$ originally defined to characterize non-stationary neuron spike trains~\cite{Lv1, Lv3, Lv5, LVcorrelation} and very recently has been used to analyze hashtag spike trains~\cite{10.1371/journal.pone.0131704, LvCollectiveAttention}. Unlike to the memory coefficient and burstiness parameter~\cite{0295-5075-81-4-48002}, $L_V$ provides a local temporal measurement, e.g. at $\tau_i$ of a successive time sequence of a spike train $\ldots$, $\tau_{i-1}$, $\tau_i$, $\tau_{i+1}$, $\ldots$, and so compares temporal variations with their local rates~\cite{Lv5}
\begin{equation}
L_V=\frac{3}{N-2}\sum\limits_{i=2}^{N-1} \left(\frac{(\tau_{i+1}-\tau_i)-(\tau_{i}-\tau_{i-1})}{(\tau_{i+1}-\tau_i)+(\tau_{i}-\tau_{i-1})}\right)^2 
\label{Eq:Lv_tau}
\end{equation}
where $N$ is the total number of spikes. Eq.~\ref{Eq:Lv_tau} also takes the form~\cite{Lv5}
\begin{equation}
L_V=\frac{3}{N-2}\sum\limits_{i=2}^{N-1} \left(\frac{\Delta\tau_{i+1}-\Delta\tau_{i}}{\Delta\tau_{i+1}+\Delta\tau_{i}}\right)^2
\label{Eq:Lv_Deltatau}
\end{equation}
Here, $\Delta\tau_{i+1}$ = $\tau_{i+1}-\tau_i$ quantifying the forward delays and $\Delta\tau_{i}$ = $\tau_{i}-\tau_{i-1}$ representing the backward waiting times for an event at $\tau_{i}$. Importantly, the denominator normalizes the quantity such as to account for local variations of the rate at which events take place. By definition, $L_V$ takes values in the interval (0:3)~\cite{10.1371/journal.pone.0131704}. It has been shown that $L_V$ classifies the salient dynamic patterns successfully~\cite{Lv1, Lv3, LVcorrelation, 10.1371/journal.pone.0131704, LvCollectiveAttention}. Following the analysis of Gamma processes~\cite{Lv1, Lv3, 10.1371/journal.pone.0131704} conventionally applied to model inter-event intervals and the neuron spike analysis~\cite{LVcorrelation}, while $L_V$ = 1 for uncorrelated (Poisson random) irregular spike trains, $L_V\approx 3$ proves that bursts dominate the spike trains and the presence of highly regular patterns in the trains gives $L_V\approx 0$. 

\begin{figure}[h!]
\begin{center}
\includegraphics[width=8.5cm]{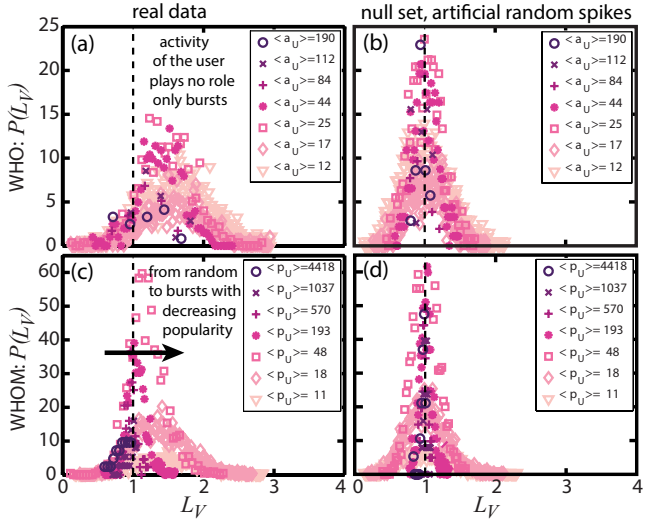}
\end{center}
\vspace{-6mm} \caption{ \label{fig:03} \small Probability density function of the local variation $L_V$, $P(L_V)$ of who (a, b) and whom (c, d) users in various ranges of the two communication frequencies, e.g. $a_U$ and $p_U$. (a, c) describe the results of the real data. When we only observe bursty communication patters in who users independent of the average user activity frequency $\langle a_U\rangle$ in (a), significant variations in $L_V$ by increasing the average user popularity $\langle p_U\rangle$ are clear in (c). The results prove that popular users in Twitter are addressed randomly in time and slightly more regular patterns observed in the most popular users. On the other hand, (b, d) present the statistics of artificially generated random spikes serving as a null model and all frequency ranges give the distributions around 1, as expected for temporarily uncorrelated signals.}
\end{figure}

We now investigate the $L_V$ analysis on the user communication spike trains. Eq.~\ref{Eq:Lv_Deltatau} is performed through the spike trains with removing multiple spikes taking place within one second. Such events are rare and their impact on the value of $L_V$ has been shown to be limited~\cite{10.1371/journal.pone.0131704}. Fig.~\ref{fig:03} describes the distribution of $L_V$, $P(L_V)$ of full spike trains all together with RT, @, and RE for the who (a, b) and whom (c, d) users. Grouping $L_V$ based on the frequency $f_U$, e.g. the activity of the who users $a_U$ and the popularity of the whom users $p_U$, we examine the temporal patterns of the trains in different classes of $a_U$ and $p_U$. For the real data in (a, c), in Fig.~\ref{fig:03}(a), $L_V$ is always larger than 1 in any values of $a_U$, suggesting that all who users contact to the whom users in bursty communications. However, in Fig.~\ref{fig:03}(c), we observe distinct behavior of the whom users and bursts present only for low $p_U$. By increasing $p_U$, $L_V\approx 1$ indicating that there is no temporal correlations among the who users referring the whom users and $L_V$ is slightly smaller than 1 for the most popular users, indicating a tendency towards regularity in the time series, as also observed for the hashtag spike trains~\cite{10.1371/journal.pone.0131704}. These observations are significantly different for artificial spike trains constructed by randomly permuting the real full spike train and so expected to generate non-stationary Poisson processes. Therefore, all distributions are centered around 1 in this case, independently of $a_U$ and $p_U$, as shown in  Figs.~\ref{fig:03}(b, d). The randomization and obtaining a null set follow the same procedure explained in detail in Ref.~\cite{10.1371/journal.pone.0131704}.

\begin{figure}[h!]
\begin{center}
\includegraphics[width=8.4cm]{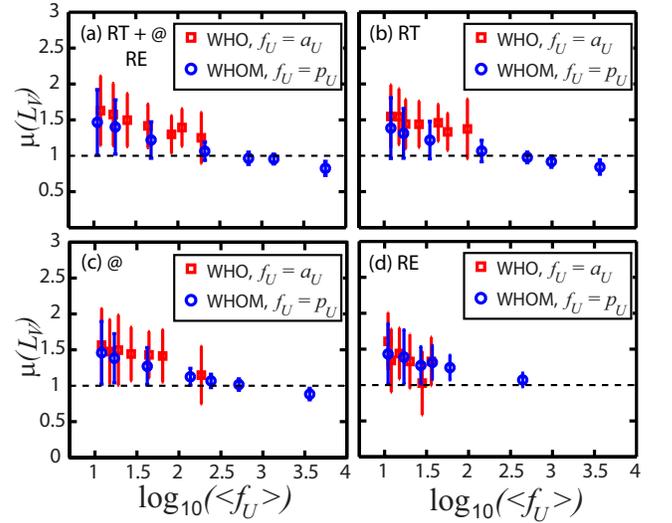}
\end{center}
\vspace{-6mm} \caption{ \label{fig:04} \small Mean $\mu$ of the local variation $L_V$ of the user communication spike trains versus the logarithmic average frequency log$_{10}\langle f\rangle$. The results of who users are represented by red squares and blue circles describe that of whom users. Types of the interaction are investigated in detail: (a) All communications of retweet, RT, mention, @, and reply, RE. (b) Only RT. (c) Only @. (d) Only RE. Independent of the types of the interaction, the frequency of communication, e.g. the activity of users $a_U$ and the popularity of users $p_U$, designs overall communication patterns. While low $f_U$ gives bursty patterns with $L_V>1$, moderate $f_U$ indicates irregular uncorrelated (Poisson random) signals, e.g. $L_V\approx1$. For all high $f_U$, $L_V<1$ presenting the regularity of the communications. The error bars show the corresponding standard variations.}
\end{figure}

Even though Fig.~\ref{fig:03} represents $P(L_V)$ of full spike trains, i.e. all interactions together, $P(L_V)$ of individual RT, @, and RE communication spike trains describes very similar temporal behavior for both the who and whom users. Fig.~\ref{fig:04} summarizes the detail of $P(L_V)$, the mean of $L_V$, $\mu(L_V)$ with the corresponding standard deviations $\sigma(L_V)$ as error bars, comparatively. The results highlight that to classify the communication temporal patterns neither the position of the users, whether active or passive, nor the types of the interaction, but the frequency of the communication $f_U$ such as $a_U$ and $p_U$ plays a major role. All Figs.~\ref{fig:04}(a-d), we observe three regions: Bursts in low $f_U$, log$_{10}\langle f_U\rangle <$ 2.5, irregular uncorrelated (Poisson random) dynamics in moderate and high $f_U$, log$_{10}\langle f_U\rangle\approx$ 2.5-3, and regular patterns in very high $f_U$, log$_{10}\langle f_U\rangle >$ 3. This conclusion supports the importance of frequency so time parameter overall human behavior~\cite{BarbarasiOriginofBursts, CorrelationCommuication}.

\begin{figure}[h!]
\begin{center}
\includegraphics[width=7cm]{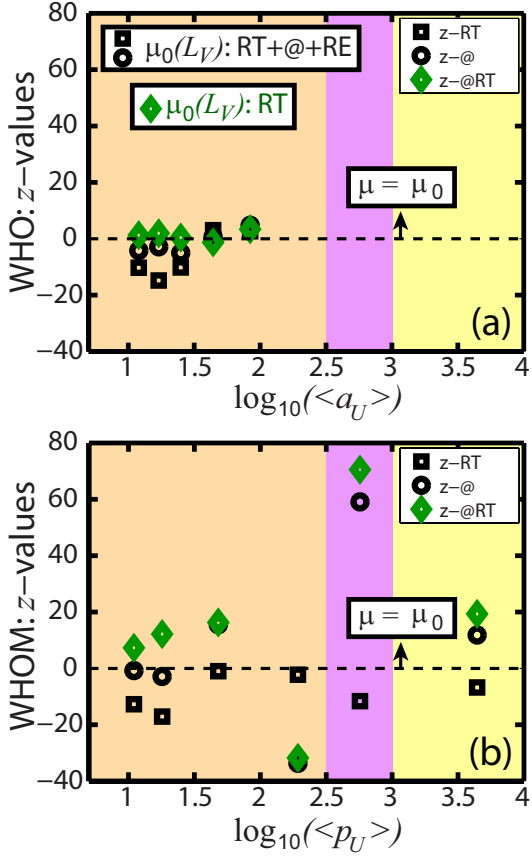}
\end{center}
\vspace{-6mm} \caption{ \label{fig:05} \small Detail comparison between the temporal patterns of different communications in each frequency range. While $x-$axis is the logarithmic average of frequency, e.g. (a) log$_{10}\langle a_U\rangle$ for the who users and (b) log$_{10}\langle p_U\rangle$ for the whom users, $y-$axis provides the calculation of three $z-$values (i) $z-$RT, the comparison of $L_V$ of the full spike train with $L_V$ of RT, in black squares, (ii) $z-$@, the same with $L_V$ of @, presented in black circles, and (iii) $z-$@RT, the comparison between $L_V$ of @ and RT, shown in green diamonds. All $z-$values are consistent with each other such that except moderate frequency range in (b), e.g. $z-$@ and $z-$@RT, we observe small $z$ concluding that the temporal patterns in the similar frequency ranges are in a good agreement. Three distinct regions are colored due to the discovered patterns in calculating $L_V$ in Fig.~\ref{fig:04}. Orange shaded area describes the ranges of the bursty patterns ($a_U$ and low $p_U$), purple area is for the irregular uncorrelated -Poisson random- patterns (moderate $p_U$), and yellow area covers the regular patterns (high $p_U$).}
\end{figure}

We now perform more detail comparison in Fig.~\ref{fig:05}, how $L_V$ of different interactions in the same frequency range varies from each other. To this end, we calculate the standard $z-$values in two ways. First, to compare $L_V$ of the full spike trains with $L_V$ of only RT and also with $L_V$ of only @ spike trains, $L_V^{\mbox{\tiny RT}}$ and $L_V^{\mbox{\tiny @}}$, respectively, we introduce 
\begin{equation}
z(f_U) = \frac{\mu(L_V^k)-\mu_0(L_V)}{\sigma(L_V^k)/\sqrt{f_U^k}}
\label{Eq:z1}
\end{equation}
Here, $k$ in superscripts labels the interaction, e.g. either RT or @. Precisely, $L_V^k$ is determined based on a filtered spike train composed of the user timestamps of either RT or @, as already used in Fig.~\ref{fig:04}(b-c). In addition, $\mu^k$ is the mean of $L_V^k$, also presented in Fig.~\ref{fig:04}(b-c), and $\mu_0$ is the mean $L_V$ of the full spike train, given in Fig.~\ref{fig:04}(a). 

In Fig.~\ref{fig:05}, black squares show $z-$values of RT and black circles describe $z-$values of @. For the who users in Fig.~\ref{fig:05}(a) where $L_V$ only presents bursty patterns (orange shaded area) and low $a_U$, we have small $z-$values proving the agreement of the temporal patterns suggested by $L_V$ in the same $a_U$. However, for the whom users in Fig.~\ref{fig:05}(b) where we have rich values of $p_U$ compared to the values of $a_U$, while $z-$values are small in bursty patterns (low $p_U$, orange area) as also observed in the who users and in regular patterns (high $p_U$, yellow area), larger $z-$@ value (the black circle) is calculated in uncorrelated Poisson dynamics (moderate $p_U$, purple area). The disagreement of $L_V$ with large $z-$@ indicates that even though $L_V\approx$ 1 in this region the results of @ are quite sensitive in the same $p_U$, which is not observed in $z-$RT (the black square).

Furthermore, we repeat the analysis across communication channels by comparing temporal patterns of RT and @ as follows 
\begin{equation}
z(f_U) = \frac{\mu(L_V^{\mbox{\tiny @}})-\mu_0(L_V^{\mbox{\tiny RT}})}{\sigma(L_V^{\mbox{\tiny @}})/\sqrt{f_U^{\mbox{\tiny @}}}}
\label{Eq:z2}
\end{equation}
The corresponding $z-$values, $z-$@RT are presented in green diamonds in Fig.~\ref{fig:05}. Comparing to the previous $z-$RT and $z$-@, we now obtain lower values for the who users [Fig.~\ref{fig:05}(a)] showing a good agreement between RT and @ patterns. Moreover, we have very similar trend for the whom users [Fig.~\ref{fig:05}(b)] as before in orange and yellow areas and large fluctuations are observed only in purple area.

\section{Correlation of $L_V$ in User Communication Habits}
\label{sec:TP}

In this final section, our interest turns into building new measures to quantify how the local variation $L_V$ fluctuates inside different classes of the frequency, $f_U$. What extend temporal communication habits of two users in the same $f_U$ ranges are dependent on each other is the first question we address. Second, we examine whether the temporal patterns of the interactions are consistent with each other for the same users and how the metric varies with increasing $f_U$.

We consider $r_{ij}^{kk^\prime}(f_U)$, the Pearson correlation coefficient of $L_V$ of two different users selected independently from the same $f_U$ classes
\begin{equation}
r_{ij}^{kk^\prime}(f_U) = \frac{\sum\limits_{i, j = 1, i\neq j}^{N_U}  [L_{V_i}^k- \mu(L_{V_i}^k)] [L_{V_j}^{k^\prime} - \mu(L_{V_j}^{k^\prime})]}{\sigma(L_{V_i}^k)\sigma(L_{V_j}^{k^\prime})}
\label{Eq:rUiUj}
\end{equation}
where $\sigma(L_{V_i}^k)=\sqrt{\sum\limits_{i=1}^{N_U} [L_{V_i}^k- \mu(L_{V_i}^k)] ^2}$. Here, $L_{V_i}$ and $L_{V_j}$ are the local variations of user $i$ and $j$, respectively, $\mu$'s are the corresponding mean values, and $N_U$ is the total number of users. Moreover, $k$ and $k^\prime$ represent all permutations among the full, RT, and @ spike trains. Furthermore, $r_{ij}^{kk^\prime}(f_U)$ is evaluated for the who and whom users, separately. Therefore, $i$ and $j$ are different users, but from the same (who/whom) pool and in the same frequency classes of $a_U$ and $p_U$, as grouped in Fig.~\ref{fig:03}. Note that before performing Eq.~\ref{Eq:rUiUj}, the corresponding $L_V$'s in the same $f_U$ class are ordered from the highest to the smallest (or vice versa) not to deform $r_{ij}^{kk^\prime}(f_U)$ artificially due to the random selection.

\begin{figure}[h!]
\begin{center}
\includegraphics[width=8.3cm]{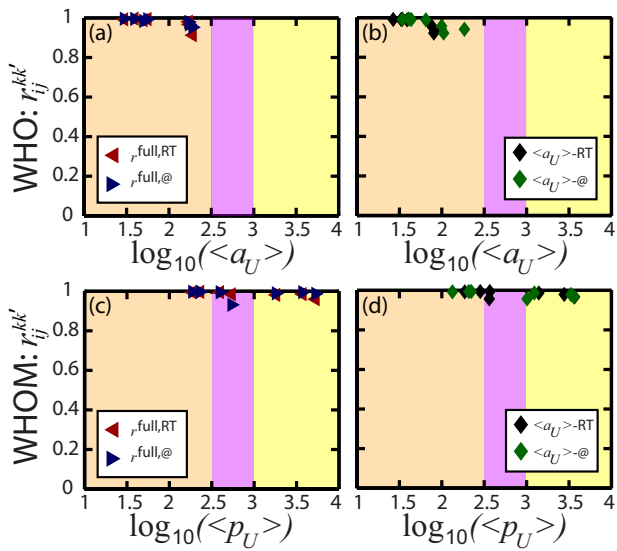}
\end{center}
\vspace{-6mm} \caption{ \label{fig:06} \small Linear correlations of $L_V$ of user pairs: The standard Pearson correlation coefficient quantifies the dependency on the temporal communication habits of two different users independently chosen from the same frequency classes, as introduced in Fig.~\ref{fig:03}. The coefficient covers 3 potential relations in the communication interactions, e.g. full and RT spike trains, red (left) triangles, full and @, blue (right) triangles, and finally RT and @, black and green diamonds. These 3 coefficients are calculated for who (a, b) and whom users (c, d), separately. 6 coefficients in total prove that the temporal patterns present high consistency in each average frequency classes, the activity $\langle a_U\rangle$ and the popularity $\langle p_U\rangle$. In (b, d), the corresponding coefficients are described with the sensitivity of the frequency classes since the average frequency in the class of RT is so similar, but not exactly equal to that of @. The colored areas are as defined in Fig.~\ref{fig:05} and characterize the three main regions of the temporal patterns of the individual user spike trains, e.g. bursts (orange), irregular random (purple), and regular patterns (yellow).}
\end{figure}

Fig.~\ref{fig:06} presents the results of $r_{ij}^{kk^\prime}(f_U)$ for the who users in (a, b) and the whom users in (c, d). Similar to $z-$values performed in the previous Section, we suggest three correlation coefficients: Red (left) triangles describe $r_{ij}^{\mbox{\footnotesize full,RT}}$, blue (right) triangles are for $r_{ij}^{\mbox{\footnotesize full,@}}$, and black and green diamonds show the values of $r_{ij}^{\mbox{\footnotesize RT,@}}$. The average frequency of the users $\langle f_U\rangle$ in the same class is similar but not equal and that is why Figs.~\ref{fig:06}(b, d) are plotted with respect to both the mean frequencies of RT and @, e.g. the average activity $\langle a_U\rangle$ and popularity $\langle p_U\rangle$ of RT and @. All correlations are above 0.85 proving the high dependency of the communication patterns of the users in the same $\langle f_U\rangle$, independent of the types of the interaction.

We now consider Eq.~\ref{Eq:rUiUj} with imposing the same user and repeat the procedure above for the correlation coefficient

\begin{equation}
r_{i}^{kk^\prime}(f_U) = \frac{\sum\limits_{i}^{N_U}  [L_{V_i}^k- \mu(L_{V_i}^k)] [L_{V_i}^{k^\prime} - \mu(L_{V_i}^{k^\prime})]}{\sigma(L_{V_i}^k)\sigma(L_{V_i}^{k^\prime})}
\label{Eq:rUiUi}
\end{equation}

\begin{figure}[h!]
\begin{center}
\includegraphics[width=8.3cm]{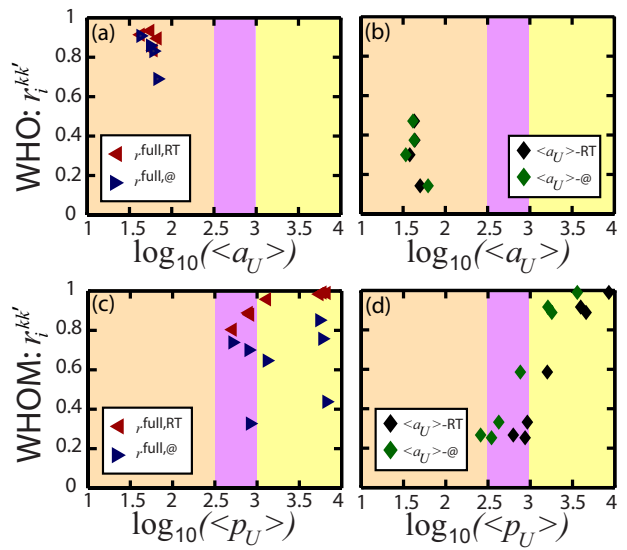}
\end{center}
\vspace{-6mm} \caption{ \label{fig:07} \small Linear correlations of $L_V$ of the \textit{same} users: The procedure and representation of the coefficients follow the same procedure as introduced in Fig.~\ref{fig:06}. However, we now impose the same users in the same frequency classes. Even though (a, c) present the agreement in the temporal patterns of full and RT spike trains of the same users, with high correlation coefficients in almost all frequency ranges, (b) indicates lower consistency between RT and @ spike trains during entire activity $\langle a_U\rangle$ and (d) provides a significant result. While less temporal coherence is observed between RT and @ spike trains in low popularity $\langle p_U\rangle$, the correlation drastically increases with $\langle p_U\rangle$.}
\end{figure}

Fig.~\ref{fig:07} summarizes the results of Eq.~\ref{Eq:rUiUi}. While Figs.~\ref{fig:07}(a, c) are in parallel with that of Fig.~\ref{fig:06} with slightly lower correlations for @ (blue right triangles), distinct behavior is observed in Figs.~\ref{fig:07}(b, d).  Low correlations in Fig.~\ref{fig:07}(b) indicate that the same who users present different temporal behavior in RT and @. On the other hand, Fig.~\ref{fig:07}(d) shows an interesting temporal habit of the whom users. Having no remarkable dependency captured in low popular users, we show that the correlation increases with $\langle p_U\rangle$ describing that the popular users are addressed in RT and @ in a temporarily similar procedure. 

\subsection{Nomenclature}
\begin{itemize}
\item OSM: Online Social Media,
\item @: Mention a user name in a tweet message,
\item RE: Reply to a tweet or retweet message,
\item RT: Retweet, share a message of other users in her/his own tweet blog,
\item WHO: Twitter users starting an interaction via @ or RE or RT with any other users,
\item WHOM: Twitter users addressed by the who users such that their message is retweeted or user name is mentioned in a message by the who users or they get a reply from the who users.\\

\noindent Any relation between who and whom users such as the following-follower is not imposed. 
\end{itemize}

\section{Discussion}
\label{sec:Discussion}

In this paper, our interest is to quantify online user communication in Twitter. To reduce the complexity in the communication, the data studied here consider only a unique subject which users talk about such as the discovery of the Higgs boson on July 4, 2012 within a restricted time window, e.g. 6 days~\cite{scientificrumor}. The main aim is to extract salient temporal patterns of communication in various types of interaction observed in Twitter such as retweet (RT), mention (@), and reply (RE). Adopting the technique so-called local variation $L_V$ originally introduced for neuron spike trains~\cite{Lv1, Lv3, Lv5, LVcorrelation} and recently has applied to hashtag spike trains in Twitter~\cite{10.1371/journal.pone.0131704, LvCollectiveAttention}, we perform detail analysis on user communication spike trains. Showing strong influences of the frequency of the hashtag spike trains on the resultant temporal patterns in the earlier work~\cite{10.1371/journal.pone.0131704, LvCollectiveAttention}, in parallel we here examine the differences in the patterns induced by the frequency of the user communication spike trains, $f_U$.

We investigate user communication spike trains in two categorizations, first set of users are the active ones, who users, and the other set is composed of the passive users, whom users, in the communication. For the who users, $f_U$ simply gives what extend the users contact to the whom users and so it is the activity of the who users, $a_U$. On the other hand, for the whom users, the generated spike trains present how often the who users refer the messages or the user names of the whom users and therefore, $f_U$ is the popularity of the whom users, $p_U$. Providing comparative statistics on $L_V$ of who and whom with increasing $a_U$ and $p_U$, respectively, we observe quite distinct temporal behavior of online users. First, we observe an asymmetry between active and passive interactions, as only the former give rise to hubs, with few users attracting a large share of the attention. Moreover, who users constantly present bursty patterns, $L_V>1$ for all values of $a_U$, whereas whom users demonstrate various dynamic behavior patterns, depending on their popularity: The least popular users with low $p_U$ experience bursty time series, popular users with moderate and high $p_U$ are contacted by temporarily uncorrelated who users and so show Poisson random spike trains $L_V\approx1$, and the most popular users with the maximum $p_U$ are referred regularly in time, e.g. $L_V<1$.

These scenarios are independent of both the position of the users, e.g. who or whom, and the preferred interactions, e.g. whether RT or @, suggesting that the frequency of the communication dominates to design social dynamic behavior. This conclusion is also supported by the high correlation coefficient of $L_V$ on the user pairs in the same frequency classes. Furthermore, the linear correlation of $L_V$ on the same users reveals interesting patterns. There, we observe that only popular users have similar dynamic behavior in both RT and @, which confirms that both metrics are complementary to characterize the influence of users.

The analysis could be specified by integrating the communication spike trains with the following-follower relation in Twitter, and focusing on the who and whom trains of connected users. An important concern is the limited time period of the data which the collection started 3 days before the announcement of the discovery and continued until 3 days after this date. Yet, it has been shown that the dynamics of the communication is drastically different before/after and during the announcement~\cite{scientificrumor}, and this variation could be investigated in our analysis. Our study shares the similar aims of the other research on online user behavior and the influence of the frequency in online platforms such as Flickr, Delicious and StumbleUpon, which user profiles have been included in the analysis~\cite{Meo:2014:AUB:2542182.2535526}. This understanding could be also applied to our analogy with considering further details in the data.

\subsection{Data Sharing}

The full data studied in this paper has open access~\cite{scientificrumor, snapnets}.

\section*{Disclosure/Conflict-of-Interest Statement}

The authors declare that the research was conducted in the absence of any commercial or financial relationships that could be construed as a potential conflict of interest.

\section*{Author Contributions}

Conceived and designed the experiments: CS. Performed the experiments: CS. Analyzed the data: CS. Contributed reagents/materials/analysis tools: RL CS. Wrote the paper: CS RL.

\section*{Acknowledgments}
C. Sanl{\i} acknowledges supports from the European Union 7th Framework OptimizR Project and FNRS (le Fonds de la Recherche Scientifique, Wallonie, Belgium). This paper presents research results of the Belgian Network DYSCO (Dynamical Systems, Control, and Optimization), funded by the Interuniversity Attraction Poles Programme, initiated by the Belgian State, Science Policy Office.

\textit{Funding.} The EU 7th Framework OptimizR Project: 48909A2 CE OPTIMIZR (Grant holder: RL, Funding receiver: CS - http://optimizr.eu/) and F.N.R.S MIS F4527.12 48888F3 (Grant holder: RL, Funding receiver: CS - http://www.fnrs.be/).

\bibliography{refs_arXiv_Frontiers_TPUsers_SanliLambiotte}

\end{document}